\documentclass[showpacs,preprintnumbers,
superscriptaddress,amsmath]{revtex4}
\usepackage[dvips]{graphicx}

\newcommand{\NPA}[3]{Nucl.\ Phys.\ {\bf A#1},\ #2 (#3)}
\newcommand{\NPB}[3]{Nucl.\ Phys.\ {\bf B#1},\ #2 (#3)}

\newcommand{\PLB}[3]{Phys.\ Lett.\ B\ {\bf #1},\ #2 (#3)}
\newcommand{\PR}[3]{Phys.\ Rep.\ {\bf #1},\ #2 (#3)}
\newcommand{\PRL}[3]{Phys.\ Rev.\ Lett.\ {\bf #1},\ #2 (#3)}

\newcommand{\PRD}[3]{Phys.\ Rev.\ D\ {\bf #1},\ #2 (#3)}
\newcommand{\PRev}[3]{Phys.\ Rev.\ {\bf #1},\ #2 (#3)}

\newcommand{\ibid}[3]{{\bf #1},\ #2 (#3)}
\newcommand\g{\gamma}
\renewcommand\d{\delta}

\newcommand\m{\mu}
\newcommand\n{\nu}

\newcommand\f{\phi}

\newcommand{\ovl}[1]{\overline{#1}}

\newcommand{\non}{\nonumber\\}

\newcommand{\be}{\begin{equation}}
\newcommand{\ee}{\end{equation}}
\newcommand{\bea}{\begin{eqnarray}}
\newcommand{\eea}{\end{eqnarray}}
\newcommand{\ba}[1]{\begin{array}{#1}}
\newcommand{\ea}{\end{array}}

\newcommand{\eqrf}[1]{Eq.\ (\ref{#1})}

\newcommand{\diracslash}[1]{#1\llap{/\kern2pt}}

\begin{document}

\title{Generalized Ward identity and gauge invariance of the 
color-superconducting gap}

\author{De-fu Hou}
\email{hou@th.physik.uni-frankfurt.de}
\affiliation{Institut f\"ur Theoretische Physik,
J.W. Goethe-Universit\"at, D-60054 Frankfurt/Main, Germany}
\affiliation{Institute of Particle Physics, Huazhong Normal University, Wuhan,
 430079, P.R.\ China}

\author{Qun Wang}
\email{qwang@th.physik.uni-frankfurt.de}
\affiliation{Institut f\"ur Theoretische Physik,
J.W. Goethe-Universit\"at, D-60054 Frankfurt/Main, Germany}
\affiliation{Physics Department, Shandong University, Jinan,
Shandong, 250100, P.R.\ China}

\author{Dirk H.\ Rischke}
\email{drischke@th.physik.uni-frankfurt.de}
\affiliation{Institut f\"ur Theoretische Physik,
J.W. Goethe-Universit\"at, D-60054 Frankfurt/Main, Germany}

\date{\today}

\begin{abstract}
We derive a generalized Ward identity for color-superconducting
quark matter via the functional integral approach. The identity implies
the gauge independence of the color-superconducting gap
parameter on the quasi-particle mass shell to subleading order in
covariant gauge.
\end{abstract}


\maketitle

Due to asymptotic freedom of quantum chromodynamics (QCD) 
the strong coupling constant $g$ becomes small at large momentum \cite{asymp}.
In cold, dense quark matter, the typical momentum scale
is of the order of the quark chemical potential $\mu$.
Consequently, at asymptotically large densities, where
$\mu \gg \Lambda_{\rm QCD}$, quark matter is a weakly coupled system.
In this case, the dominant interaction between two quarks
is single-gluon exchange, which is attractive in the
color-antitriplet channel. Due to Cooper's theorem \cite{bcs}, 
at sufficiently low temperatures the quark Fermi surface becomes unstable 
with respect to the formation of Cooper pairs. 
Since this is analogous to what happens
in ordinary superconductors \cite{bcs}, this phenomenon was
termed {\em color\/} superconductivity \cite{bailin}. 
Recently, color superconductivity has been extensively 
studied within QCD at weak coupling
\cite{ris03,son99,pis00,ris01,miransky}, 
in phenomenological Nambu--Jona-Lasinio-type models 
\cite{alford1,Neumann}, and in  effective theories
\cite{son00,hong-eff,cas99,schafer,ginzburg-landau}.

In a superconductor, exciting particle-hole pairs costs
at least an energy amount $2\phi _0$, where $\phi _0$
is the value of the superconducting gap parameter at the
Fermi surface for zero temperature. The gap parameter can be computed from
a gap equation derived in mean-field approximation
which involves single-gluon exchange and
bare quark-quark-gluon vertices. Schematically, this
gap equation can be written in the form \cite{pis00}
\be
\f_0 = g^2 \, \left[\; \zeta \,
\ln^2 \left( \frac{\m}{\f_0} \right)
+  \beta \, \ln \left( \frac{\m}{\f_0} \right) +  \alpha\;
\right] \; \f_0\;.
\label{gapequation}
\ee
In weak coupling, $g \ll 1$, the solution is
\cite{SchaferWilczek,pis00,son99,miransky}
\be
\f_0 = 2 \, b \, \m \, \exp \left( - \frac{c}{g} \right)
\left[ 1 + O(g) \right]\;.
\label{gapsolution}
\ee
The first term in \eqrf{gapequation} contains two powers of the logarithm
$\ln (\m/\f_0)$: one is the same as in BCS theory \cite{bcs} (the
so-called ``BCS logarithm'')
and the other arises from the collinear exchange of
almost static magnetic gluons, which is a long-range
interaction \cite{pis00,son99}.
The weak-coupling solution (\ref{gapsolution})
implies that this term contributes to the gap equation at order $O(1)$.
We call this term the leading-order term.
The value of the coefficient $\zeta$ determines
the constant $c$ in \eqrf{gapsolution}.  
The second term in \eqrf{gapequation} contains subleading
contributions of order $O(g)$ to the gap equation, characterized by a
single power of the logarithm $\ln (\m/\f_0) \sim 1/g$.
A part of them arises from the exchange 
of non-static magnetic and static electric
gluons \cite{pis00}. Another part is due to the contribution from
the regular quark self-energy \cite{wang,bro00}. 
Vertex corrections were reported not to contribute to subleading order
\cite{bro00,mishra}. The coefficient $\beta$
in \eqrf{gapequation} determines the constant
$b$ in \eqrf{gapsolution}. 
The third term in \eqrf{gapequation} summarizes sub-subleading
contributions of order $O(g^2)$ with neither a collinear
nor a BCS logarithm. 

In principle, on the quasi-particle mass shell
the gap parameter is an observable quantity,
and thus must be independent of the choice of gauge.
However, the mean-field approximation to the QCD gap equation may violate this
requirement. It was argued in Refs.\
\cite{SchaferWilczek,pis00,ShusterRajagopal} that, in mean-field approximation,
gauge-dependent terms enter the QCD gap equation 
at sub-subleading order. The authors of
Ref.\ \cite{pis02} confirmed that this is indeed the case for
Coulomb gauge and when taking the gap parameter on the
quasi-particle mass shell.
In contrast to this result, it was pointed out  \cite{miransky,hong03}
that, in covariant gauge, the gauge-parameter dependence already shows 
up at subleading order. Denoting the gauge
parameter in general covariant gauges by $\xi$, there is then
an additional factor $\exp(3 \xi/2)$ to the prefactor $b$ in
\eqrf{gapsolution}.

It is a priori not clear why it should depend on the choice of gauge,
at which order gauge-parameter dependent terms appear in the
QCD gap equation in mean-field approximation.
A posteriori, this result
is maybe not that puzzling at all, since the mean-field approximation
corresponds to a resummation of a particular class of diagrams which
may contribute at different orders for different choices of gauge.
In any case, in order to remove the gauge-parameter dependence 
at subleading order in covariant gauge, one has to take into account 
corrections to the mean-field approximation. 
Gauge-parameter dependent terms will then only occur at sub-subleading order. 
The obvious correction so far not taken into account in the
mean-field approximation is to replace bare $qqg$ 
vertices by full vertices. It is the purpose of this paper
to show how the inclusion of vertex corrections guarantees the
gauge-parameter independence at subleading order in covariant gauge.

As a matter of fact, it is not even necessary to compute the vertex
corrections explicitly.
We shall prove the gauge independence in a rather convenient way by making use 
of a Ward identity which relates the vertex to the inverse propagator.
This approach has been frequently applied to show the gauge independence of 
physical collective excitations in thermal gauge theories, like hot QCD
\cite{rebhan,Pb}. However, in a superconductor, the existence
of a fermion-fermion condensate necessitates the use
of the Nambu-Gor'kov (NG) basis to describe the
propagation of quasi-particle excitations. It is
therefore desirable to derive the Ward identity
in the NG basis. In this paper, we 
derive this generalized Ward identity for color-superconducting 
quark matter and apply it to the gap equation, thereby showing that
the gap parameter is gauge-independent to subleading order on
the quasi-particle mass shell.

We note that our approach is similar in spirit to that of
Gerhold and Rebhan \cite{gerhold}, who
used generalized Nielsen identities to give a formal proof that 
the fermionic quasi-particle dispersion 
relations in a color superconductor are gauge independent, assuming
that the 1PI part of the variation of the 
effective action with respect to the gauge parameter
has no singularities coinciding with those of the quark propagator. 
Other Ward identities for color-superconducting quark matter
have been derived by Miransky, Shovkovy, and Wijewardhana
\cite{igor}, but these identities were not
suitable to see the gauge independence of the gap parameter 
to subleading order.

In order to derive the generalized Ward identity,
consider QCD with $N_f$ quark flavors 
and quark chemical potential $\mu$.
The Lagrangian and the generating functional 
can be written in the NG basis,
\bea
\mathcal{L}&=& \frac{1}{2}\, \ovl{\Psi}\mathcal{S}_0^{-1}\Psi
+\frac{g}{2}\,\ovl{\Psi}\mathcal{T}^a\g_\m \Psi A_a^\m 
-\frac{1}{4}\,F^{a}_{\m\n}\, F^{\m\n}_a
-\frac{1}{2\xi}(\partial_\m A_a^\mu)^2 \; , \label{L} \\
Z[J,\ovl{H},H]
&=&\int [dA][d\ovl{\Psi}][d\Psi]
\exp \left\{ \int_X\, \left[ \mathcal{L}
+J_\mu^a A_a^\mu+\ovl{H}\Psi +\ovl{\Psi}H \right] 
\right\}\,\, .
\label{genfuncquarks3}
\eea
The quark fields and their sources are given by
\be
\Psi = \left( \begin{array}{c} 
                    \psi \\
                    \psi_C 
                   \end{array}
            \right) \, , \,\,\,\,
\ovl{\Psi} = ( \ovl{\psi} \, , \, \ovl{\psi}_C ) , \,\,\,\,
H = \left( \begin{array}{c} 
                    \eta \\
                    \eta_C 
                   \end{array}
            \right) , \,\,\,\,
\ovl{H} = ( \ovl{\eta} , \, \ovl{\eta}_C )\, , 
\ee  
where the charge conjugate spinors 
$\psi_C,\, \ovl{\psi}_C$ are defined through
$\psi _C=C\ovl{\psi}^T$, $\ovl{\psi}_C=\psi ^TC$
with the charge conjugation matrix $C=i \gamma^2 \gamma_0$.
The inverse free quark propagator (including the chemical potential
$\mu$) is 
\be
 \mathcal{S}_0^{-1}\equiv
\left(\ba{cc} {S}_{0,11}^{-1}  & 0\\ 
              0 & {S}_{0,22}^{-1}  \ea \right)\;,
\ee 
where $S_{0,11}^{-1} = i\gamma^\mu \partial_\mu + \mu \gamma_0 - m$,
$S_{0,22}^{-1} = i\gamma^\mu \partial_\mu - \mu \gamma_0 - m$.
The color part of the $qqg$ vertex is
\be
\label{Gamma}
\mathcal{T}^a 
\equiv \left( \begin{array}{cc}
T^a & 0      \\
0   & -T^{aT}
\end{array} \right)\; ,
\ee
where $T^a$ are the generators of $SU(3)_c$.
The factor $1/2$ in front of the first two terms in \eqrf{L} accounts
for the doubling of the quark degrees of freedom in the NG basis.
The gluon field and the field strength tensor
are denoted by $A^a_\m$ and $F^a_{\m\n}$, respectively. 
The last term in $\mathcal{L}$ 
is the gauge fixing term, and $J_\mu^a$ is the source term
for the gluon field.
We denote the space-time integration 
as $\int_X \equiv \int_0^{1/T} d\tau \int_V d^3{\bf x}\,$. 
We neglect ghost terms, as ghosts 
do not contribute to
leading and subleading order in
cold, dense quark matter \cite{ris03}.

The generating functional $Z$ 
is invariant under the following infinitesimal 
$SU(3)_c$ gauge transformation 
\bea
\delta\Psi&=&i\theta^a \mathcal{T}^a\Psi\;, \non
\delta\ovl{\Psi}&=&-i\theta^a \ovl\Psi \mathcal{T}^a\;, \\
\delta A_\mu^a&=&\frac{1}{g}\partial_\mu 
\theta^a+f^{abc}A_\mu^b \theta^c \;. \nonumber
\eea
The variation of $Z$ with respect 
to the infinitesimal parameter $\theta^a$ of $SU(3)_c$ gauge
transformations vanishes, which yields the following identity:
\be
-\frac{1}{\xi}\left\{  
\partial^2\partial_\m \langle A^\mu_a\rangle
+g f^{abc} \left[ \frac{}{}\! \partial_\m \partial_\n D^{\m\n}_{bc}(X,X)+ 
\langle A^\mu_b \rangle \partial _\m \partial _\n
\langle A^\nu_c\rangle \right] \right\}
- \partial_\mu J^\mu_a
+g f^{abc}J^b_\mu\langle A_c^\mu \rangle
+ig \left[ \ovl{H}\mathcal{T}^a \langle \Psi \rangle
-\langle \ovl\Psi \rangle \mathcal{T}^a H\right] =0\;,
\label{identity1}
\ee
where the average is taken in the functional sense: 
\be
\langle F(A,\ovl\Psi,\Psi) \rangle 
=\frac{1}{Z}
\int [dA][d\ovl\Psi][d\Psi]\; F(A,\ovl\Psi,\Psi)\; 
\exp \left\{ \int_X\, \left[ \mathcal{L}
+J_\mu^a A_a^\mu+\ovl{H}\Psi +\ovl{\Psi}H \right] 
\right\} \;,
\label{identity}
\ee
and $D^{\m\n}_{bc}=\d ^2W/\d J_\m ^b\d J_\n ^c$ is the
two-point Green's function, where $W=\ln Z$ is the 
generating functional for connected Green's functions. 
In the hard-dense-loop (HDL) approximation,
the two-point function $D^{\m\n}_{bc}$ is symmetric in the color
indices $b$ and $c$. Consequently, the term proportional 
to $D^{\m\n}_{bc}$ in \eqrf{identity1} vanishes,
because the structure constants $f^{abc}$ multiplying the
two-point function are antisymmetric. In a color
superconductor this does not need to be true \cite{ris01}, but
the violation of symmetry occurs for gluon
energies and momenta of order $\sim \phi_0$.
In the QCD gap equation, this range of gluon energy and momentum
contributes only beyond subleading order \cite{ris01}.

We can use the effective action 
$\Gamma[A,\ovl{\Psi},\Psi] 
\equiv W-\int _X [J_\mu^a A_a^\mu+\ovl{H}\Psi +\ovl{\Psi}H ]$ to 
rewrite the identity (\ref{identity1}), i.e., 
we replace $J_a^\mu$ by $- \d \Gamma/ \d A^a_\mu$, $\ovl{H}$ by
$\d\Gamma/ \d\Psi$, and $H$ by $- \d \Gamma/\d \ovl{\Psi}$. 
Note that, from now on,
$A_\mu^a,\, \ovl{\Psi}$, and $\Psi$ denote the 
{\em expectation values\/} of the gluon and quark fields.
Taking the functional derivative $\d /\d \Psi (X_1)$ 
from the right and $\delta /\delta \ovl\Psi (X_2)$ from the left, 
and using the fact
that the expectation values of the Grassmann-valued 
quark fields always vanish, $\ovl{\Psi} = \Psi \equiv 0$,
we obtain the following identity 
\bea
- \frac{i}{g} \left[ \partial_\m ^X \delta^{ab} - g f^{abc} A^c_\mu(X) \right]
\frac{\,\delta^3\, \Gamma[A,0,0]}{\delta \ovl\Psi(X_1)\delta A^b_\mu(X) 
\delta \Psi(X_2)}
& = & \delta^{(4)}(X-X_2)  \frac{\delta^2\, \Gamma[A,0,0]}
{\delta \ovl\Psi(X_1)  \delta \Psi(X_2) }\,\mathcal{T}^a  \non
&  &-  \delta^{(4)}(X-X_1)\mathcal{T}^a \frac{\delta^2\, \Gamma[A,0,0]}
{\delta \ovl\Psi(X_1)  \delta \Psi(X_2) } \; . \label{identity2}
\eea
The expectation value $A_\mu^c$ of the gluon field on the left-hand
side is not necessarily zero in a color superconductor 
\cite{gerhold,kryjevski,ris}. However, 
in a two-flavor color superconductor, its value is
of order $\phi_0^2/(g^2 \mu)$ \cite{ris}, i.e., much smaller
than the derivative $\partial_\mu^X$ which
is of the order of the gluon energy, $p_0 \sim \phi_0$, or even
of the gluon momentum, $p \sim (g^2 \mu^2 \phi_0)^{1/3}$ \cite{pis00}.
Therefore, in the following we may safely neglect the term proportional to 
$A^c_\m$ on the left-hand side of \eqrf{identity2}.
We now rewrite \eqrf{identity2} in momentum space,
\be
P_\mu\Gamma^\mu_a (K,P,K+P)
= g [\mathcal{T}^a \mathcal{S}^{-1}(K+P)-\mathcal{S}^{-1}(K) 
\mathcal{T}^a] \; ,
\label{gwi}
\ee
where $\Gamma^\mu_a (K,P,K+P)$ is the Fourier transform of 
$\d ^3\Gamma/\delta \ovl\Psi(X_1)\delta A^a_\mu(X)\delta \Psi(X_2)$,
representing the full $qqg$ vertex, while
$\mathcal{S}^{-1}(K)$ is the Fourier transform of 
$\d ^2\Gamma/\delta \ovl\Psi(X_1)\delta \Psi(X_2)$, i.e., the
full inverse quark propagator.
We note that the above identity is a $2\times 2$ matrix identity 
in the NG basis, and thus a {\em generalized\/} Ward identity. 
Equation (\ref{gwi}) is rather similar to the Ward identity derived
by Nambu to prove the gauge invariance of the
Meissner effect in ordinary superconductors \cite{nambu}, 
except that $\mathcal{T}^a$ assumes the role of the Pauli matrix
$\tau_3$ in NG space.

The inverse of the quark propagator $\mathcal{S}^{-1}$ in 
Eq.\ (\ref{gwi}) is defined as \cite{ris03}
\be
\mathcal{S}^{-1}\equiv \mathcal{S}_0^{-1} + \Sigma\;, 
\label{S1-2}
\ee
where $\Sigma$ is the quark self-energy in the NG basis.
In momentum space, the Dyson-Schwinger equation for
the quark self-energy reads \cite{ris03}
%%%%%%%%%%%%%%%%%%%%%%%%%%%%%%%%%%%%
\begin{figure}
\includegraphics[scale=0.4]{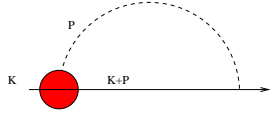}
\caption{The quark self-energy in the Dyson-Schwinger equation. 
The dashed line represents the hard-dense-loop (HDL) dressed propagator, 
the solid line is the quasi-quark propagator, and 
the black blob is the full $qqg$ vertex.}
\label{fig1}
\end{figure}
%%%%%%%%%%%%%%%%%%%%%%%%%%%%%%%%%%%%
\be
\Sigma(K)  = - g \, 
\int_P   \Gamma_\mu^a \, \mathcal{S}(K+P) \, 
\mathcal {T}^b \gamma_\nu \, D^{\mu\nu}_{ab}(P)\;.
\label{PiPscf}
\ee
The right-hand side of this equation is shown diagrammatically in
Fig.\ \ref{fig1}.
Here, $D^{\mu\nu}_{ab}$ is the HDL-resummed gluon propagator
and $\int_P \equiv T \sum_n \int d^3 {\bf p}/{(2 \pi)^3}$, where
the sum runs over all fermionic Matsubara frequencies $\omega_n
= (2 n+1) \pi T$.
The first vertex in the above equation is the full one and 
the second vertex is the bare one. 

We now show with the help of the generalized Ward identity (\ref{gwi})
that, on the quasi-particle mass shell, 
the quark self-energy (\ref{PiPscf}) does not
depend on the gauge parameter $\xi$ in covariant gauge.
To this end, consider the contribution from the gauge-dependent
part of the gluon propagator to the quark self-energy,
\be \label{sigmaxi}
\Sigma_\xi(K)  \equiv  g \xi \,
\int_P   \Gamma_\mu^a \, \mathcal{S}(K+P) \, 
\mathcal {T}^a \gamma_\nu  \frac{P^\mu\,P^\nu}{(P^2)^2}\;.
\ee
Using the generalized Ward identity (\ref{gwi}), this becomes
\be \label{sigmaxi_2}
\Sigma_\xi(K)  =  g^2 \xi \, 
\int_P \left[ \mathcal{T}^a \mathcal{S}^{-1}(K+P) -
\mathcal{S}^{-1}(K) \mathcal{T}^a \right]\, \mathcal{S}(K+P) \, 
\mathcal {T}^a \gamma_\nu  \frac{P^\nu}{(P^2)^2}\;.
\ee
Now we put the external momentum $K$ on the quasi-particle mass shell, 
which is defined as
\be \label{onshell}
S^{-1}(K)\Psi_{\rm on-shell}= 0\;,\;\;\;\; 
\ovl\Psi_{\rm on-shell}S^{-1} (K) = 0 \;,
\ee 
where $\Psi_{\rm on-shell}$ is the on-shell quasi-particle wave 
function. Then, when sandwiching \eqrf{sigmaxi_2} 
between on-shell wave functions, the second term in brackets 
in \eqrf{sigmaxi_2} vanishes. We arrive at
\be
\Sigma_\xi (K) 
= g^2 \xi \, \frac{4}{3}\, \mathbf{1}
\int_P  \frac{\gamma_\nu P^\nu}{(P^2)^2}
= 0\;,
\label{vanish}
\ee
where $\mathbf{1}$ is the $2\times2$ unit matrix in the NG 
basis. The right-hand side vanishes because
the integrand of the $P$-integral
is an odd function of $P$. Hence, we see that 
the contribution from the gauge-dependent
part of the self-energy (\ref{PiPscf})  vanishes 
on the quasi-particle mass shell. Therefore, the gap parameter
as well as the quasi-particle dispersion relation 
are gauge-independent in a color superconductor. 

This result holds strictly speaking only to subleading order 
in the QCD gap equation.
The reason is that the generalized 
Ward identity (\ref{gwi}) is not an exact 
identity, but only holds under the following
approximations. The first approximation 
is that we neglected the contribution of ghosts. 
For QCD, which is a non-Abelian gauge theory, in general one has to  
take into account ghosts and use the BRST transformation and 
the Slavnov-Taylor identity or, equivalently, generalized Nielsen
identities \cite{gerhold,rebhan}. 
However, in dense quark matter, diagrams
involving ghosts can be neglected at least to subleading order
in the QCD gap equation \cite{ris03}. 
The second approximation
is that we neglected the contribution from $D^{\m\n}_{bc}$ in
Eq.\ (\ref{identity1}), arguing that the gluon propagator 
is symmetric in the color indices. This
is true for an HDL-dressed propagator, but not necessarily for
the gluon propagator in a color superconductor \cite{ris01}.
However, terms which violate the symmetry of $D^{\mu\nu}_{bc}$
should also be negligible up to subleading order in
the gap equation.
Finally, the third approximation was to
neglect the expectation value of the gluon field
in the generalized Ward identity (\ref{gwi}).
In Ref.\ \cite{ris} it is shown that actually
$A_\mu^a \neq 0$ in order ensure color neutrality.
However, at least in a two-flavor color superconductor,
$A_\mu^a \sim \phi_0^2/(g^2 \mu)$, i.e., the expectation value
of the gluon field is negligible up
to subleading order in the gap equation.

In summary, we derived a generalized Ward identity from QCD for 
dense, color-superconducting quark matter.
The identity implies that, on the quasi-particle mass shell, 
the gap function and the 
quasi-particle dispersion relation are independent of the gauge 
parameter in covariant gauge up to subleading order.
There is one potential caveat to this statement.
What we have shown is that, to subleading order,
the gauge dependence of the quark self-energy (\ref{PiPscf}) arising
from the {\em gauge-dependent part\/} of the {\em gluon propagator\/} vanishes
on the mass shell.
In principle, however, other gauge-dependent terms arise 
from the gauge dependence of the {\em full vertex\/} $\Gamma_\mu^a$
when combined with the {\em physical part\/} of the gluon propagator
in Eq.\ (\ref{PiPscf}).
In order to show that this is of sub-subleading order
in the gap equation, one actually
has to compute the gauge-dependent part of the vertex explicitly.
This calculation can be simplified considerably 
with the help of the Ward identity
(\ref{gwi}), for more details see Ref.\ \cite{hwr}. The result is
that also this contribution is of sub-subleading order in the gap
equation.

Our result shows that in order to obtain a gauge-independent gap
function up to subleading order, one has to use the full vertex 
as well as the full fermion propagator in the NG basis. 
A consequence is that the prefactor $\exp(3 \xi/2)$ to the gap
function found in the mean-field approximation \cite{miransky,hong03}
will be removed by contributions from the full $qqg$ vertex
when taking the gap function on the quasi-particle mass shell.
An explicit diagrammatic proof of this statement will
be presented elsewhere \cite{hwr}.

\acknowledgments

The authors thank R.\ Pisarski, A.\ Rebhan,
H.-c.\ Ren, and  I.\ Shovkovy for  interesting  discussions
and a critical reading of the manuscript. 
D.\ Hou\ acknowledges financial support from the
Alexander von Humboldt-Foundation and the National Natural
Science Foundation of China under grants 1000502 and 10135030.
He also appreciates help and support from 
the Institut f\"ur Theoretische Physik of the 
J.W.\ Goethe-Universit\"at, and especially from 
Prof.\ W.\ Greiner. The work of 
Q.\ Wang is supported by GSI Darmstadt and BMBF.

\end{document}